# Structurally and Chemically Compatible BiInSe$_3$ Substrate for Topological Insulator Thin Films

Xiong Yao[1], Jisoo Moon[2], Sang-Wook Cheong[1], and Seongshik Oh[1](✉)

[1] Center for Quantum Materials Synthesis and Department of Physics & Astronomy, Rutgers, The State University of New Jersey, Piscataway, New Jersey 08854, United States
[2] Department of Physics & Astronomy, Rutgers, The State University of New Jersey, Piscataway, New Jersey 08854, United States

**ABSTRACT**
Quality of epitaxial films strongly depends on their structural and chemical match with the substrates: the more closely they match, the better the film quality is. Topological insulators (TI) such as Bi$_2$Se$_3$ thin films are of no exception. However, there do not exist commercial substrates that match with TI films both structurally and chemically, at the level commonly available for other electronic materials. Here, we introduce BiInSe$_3$ bulk crystal as the best substrate for Bi$_2$Se$_3$ thin films. These films exhibit superior surface morphology, lower defect density and higher Hall mobility than those on other substrates, due to structural and chemical match provided by the BiInSe$_3$ substrate. BiInSe$_3$ substrate could accelerate the advance of TI research and applications.

## 1 Introduction

Over the past few years, TI thin films prompted the discovery of numerous novel quantum phenomena, such as quantum anomalous Hall effect [1-3], non-conventional quantum Hall effects [4-11], axion insulator state [12-14] and quantized Faraday and Kerr rotation [15]. All these quantum effects are sensitive to disorders and require either extremely low temperatures to freeze the disorder effect or ultralow defect densities. As demonstrated before [7-10], substrate is a significant source of disorder and existing commercial substrates always lead to high level of interfacial defects due to combination of structural and chemical mismatch. Figure 1(a) shows commonly used commercial substrates and their in-plane lattice parameters in comparison with TI materials. These substrates have either entirely different crystal structures and ionic configurations (such as InP(111) and SrTiO$_3$(111)) or vastly different lattice parameters (such as Al$_2$O$_3$). These structural and/or chemical mismatch inevitably leads to high level of interfacial defects and residual carrier densities, making it difficult to access the quantum signatures of topological surface states. An ideal substrate for a TI film should have the same crystal structure with similar lattice parameter and chemical composition to the TI material.

The importance of matched substrate is well known in semiconductor devices. In general, the best substrate for any material should be the same material. For example, the best substrate for Si film should be Si substrate and the best for GaAs film should be GaAs substrate, etc. However, not every material has large enough bulk crystals that can work as substrates, and GaN, the key material for blue LED device, is a good example. Because GaN substrate does not exist, growth of GaN thin films is done on other commonly available substrates such as Al$_2$O$_3$ or SiC. However, due to significant lattice mismatch, GaN films always have high density of defects, and thick buffer layers of GaN and/or AlN are required to reduce these defects in the active region to an acceptable level [16-19]. TI films such as Bi$_2$Se$_3$ grown on commercial substrates have similar problems of high density of interfacial defects. To make matters worse, unlike GaN, interfacial defects on TI films cannot be buried because the topologically protected surface band drives these defects electronically active. Fortunately, we have previously demonstrated that chemically and structurally well matched BiInSe$_3$ buffer layers can drastically suppress these interfacial defects in Bi$_2$Se$_3$ thin films [7, 8]. However, this buffer layer scheme requires a time-consuming and complex fabrication process, hindering broad applications of TI materials. Here, we show that BiInSe$_3$ bulk crystals can negate the need for the complex buffer layer scheme and provide excellent platform for TI thin films.

## 2  Results and discussion

Bi$_2$Se$_3$, In$_2$Se$_3$ and the solid solution BiInSe$_3$ all share the same layered hexagonal structure with the weak Van der Waals bonding between QLs (quintuple layer; 1 QL is approximately 1 nm thick). Owing to the large band gap of In$_2$Se$_3$, ~1.3 eV, compared with ~0.3 eV for Bi$_2$Se$_3$, the solid solution BiInSe$_3$ not only provides ideal structural and chemical compatibility for Bi$_2$Se$_3$ growth, but also preserves the large band gap which is prerequisite to be used as substrate.

We grew the BiInSe$_3$ bulk crystal substrates by a modified Bridgman method. Starting materials of Bi (99.997%, Alfa Aesar), In (99.99%, Alfa Aesar) and Se (99.999%, Alfa Aesar) with the molar ratio of 1:1:3 were mixed together and then sealed into an evacuated conical-shaped quartz tube. The quartz tube was put into a furnace with vertical temperature gradient and held at 850 ℃ for 48h, then cooled down to room temperature to get polycrystalline BiInSe$_3$. Then the polycrystalline BiInSe$_3$ together with the quartz tube was taken out, sealed into a larger quartz tube, then put into the same furnace and heated to 950 ℃, with a subsequent slow cooling to 550 ℃ over 400 h, and cooled down to room temperature to get single crystal BiInSe$_3$. The obtained single crystals exhibit a shiny ingot with sharp conical tip, as shown in Fig. 1(b). The crystal boule can be easily cleaved by a blade along the growth direction and further cleaved by Scotch tape. The cleaved fresh BiInSe$_3$ substrate shows clean and flat surface as shown in Fig. 1(c). We performed X-ray diffraction (XRD) measurement to check the crystal quality, the result is shown in Fig. 1(d). The sharp (00l) diffraction peaks indicate good single crystalline quality of the BiInSe$_3$ substrates.

Surface flatness is one of the key parameters essential for high quality epitaxial growth [20]. Figure 1(e) shows the surface roughness on a cleaved BiInSe$_3$ substrate, as measured by atomic force microscopy (AFM): its surface roughness is 59 pm. In comparison, a common commercial Al$_2$O$_3$ substrate in Fig. 1(f) exhibits a roughness of 132 pm. The much smoother surface of the cleaved BiInSe$_3$ substrate is due to the Van der Waals nature of the Se-Se bonding: it always cleaves between the atomically-defined Se-Se layer. On the other hand, without such natural cleavage plane, commercial substrates like Al$_2$O$_3$, SrTiO$_3$ or GaAs cannot avoid a certain level of atomic-level roughness despite elaborated processes of chemical and mechanical polishing.

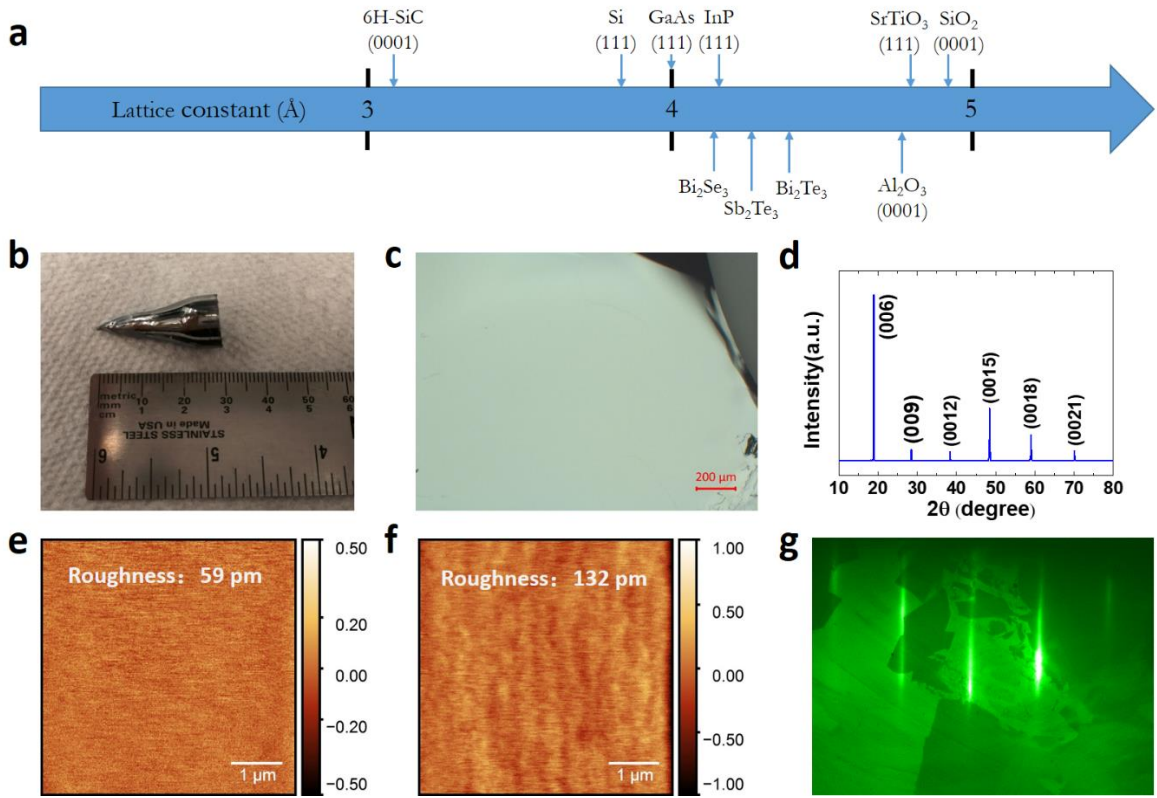

Figure 1: High quality single crystal substrate BiInSe$_3$. (a) Schematic diagram of in-plane lattice parameters for some representative commercial substrates and 3D TI materials (Bi$_2$Se$_3$, Bi$_2$Te$_3$ and Sb$_2$Te$_3$). (b) Image of the as-grown shiny BiInSe$_3$ ingot with conical tip. (c) Optical microscope image of cleaved fresh surface of a BiInSe$_3$ single crystal, exhibiting clean flat surface. (d) X-ray diffraction result of a typical BiInSe$_3$ substrate. (e, f) AFM image (5 μm × 5 μm) of a BiInSe$_3$ and commercial Al$_2$O$_3$ substrate surface, respectively. Inset words indicate the root mean square roughness of the surface. Units in the AFM scale bars are nm. (g) In situ RHEED pattern of the cleaved BiInSe$_3$ substrate surface.

For film growth, we mounted the BiInSe$_3$ substrate onto a dummy Al$_2$O$_3$ substrate, which acts as sample holder, using PELCO® high performance ceramic adhesive (Ted Pella), then cured the adhesive, cleaved a fresh surface and put it into the MBE chamber right away: more details are included in a previous report [21]. Figure 1(g) gives the reflection high-energy electron diffraction (RHEED) pattern of the BiInSe$_3$ substrate, exhibiting sharp bright streaks, which is another signature of atomically flat surface. Both AFM and RHEED measurements confirm the atomic flatness of the cleaved BiInSe$_3$ surface.

In order to get rid of any contaminants, we outgas the BiInSe$_3$ substrate at 600 ℃ for 30 min with Se flux supplied in the MBE chamber. Then after the substrate is cooled to 350 ℃, we grow Bi$_2$Se$_3$ films on the BiInSe$_3$ substrate by co-evaporating high-purity Bi and Se using standard effusion cells. Figure 2(a) shows the RHEED pattern after deposition of 10 QL Bi$_2$Se$_3$ on a BiInSe$_3$ substrate. The bright and sharp RHEED pattern indicates high quality epitaxial growth of the Bi$_2$Se$_3$ film.

Surface morphology is an important indicator of the quality of a film. Figure 2(b) gives the AFM image of a 10 QL Bi$_2$Se$_3$ film surface grown on a BiInSe$_3$ substrate. The most noticeable feature is the large flat terraces, which are much larger than those of previous Bi$_2$Se$_3$ thin films [22-28]. For comparison, we present morphologies of two control samples in Fig. 2(c) and (d). Figure 2(c) shows the AFM image of a 10 QL $Bi_2Se_3$ film directly grown on $Al_2O_3$ substrate. The surface shows characteristic triangular shaped terraces: $Bi_2Se_3$ films grown on InP(111) or Si(111) substrates exhibit even smaller terraces [26, 29]. Figure 2(d) shows the surface morphology of a 10 QL $Bi_2Se_3$ film grown on $In_2Se_3$-$BiInSe_3$ buffer layer [7], which exhibits slightly larger terraces with sharper edges than Fig. 2(c) ($Al_2O_3$ substrate), but still much smaller than Fig. 2(b) ($BiInSe_3$ substrate). Through this comparison, we can see that the $BiInSe_3$ substrate is clearly better than both $Al_2O_3$ substrate and $In_2Se_3$-$BiInSe_3$ buffer layer in terms of surface morphology. The lower $Bi_2Se_3$ nucleation density (thus larger terrace size) on $BiInSe_3$ suggests lower defect density in the film, which provides an evidence for the higher crystalline quality. Moreover, because the $In_2Se_3$-$BiInSe_3$ buffer layer is itself grown on ill-matched $Al_2O_3$ substrate, it cannot avoid these relatively small triangular terraces on its own. On the other hand, $BiInSe_3$ substrate is a single crystal grown by Bridgeman method, and can be, after cleavage, atomically flat, free of terraces over a macroscopic scale as shown in Fig. 1(e). This large flat surface of well-matched substrate provides the ideal platform for $Bi_2Se_3$ thin film growth as shown in Fig. 2(b). As we can see from the following part, the superior surface morphology of $Bi_2Se_3$ film grown on $BiInSe_3$ substrate also plays a very critical role in improving the electrical transport performance.

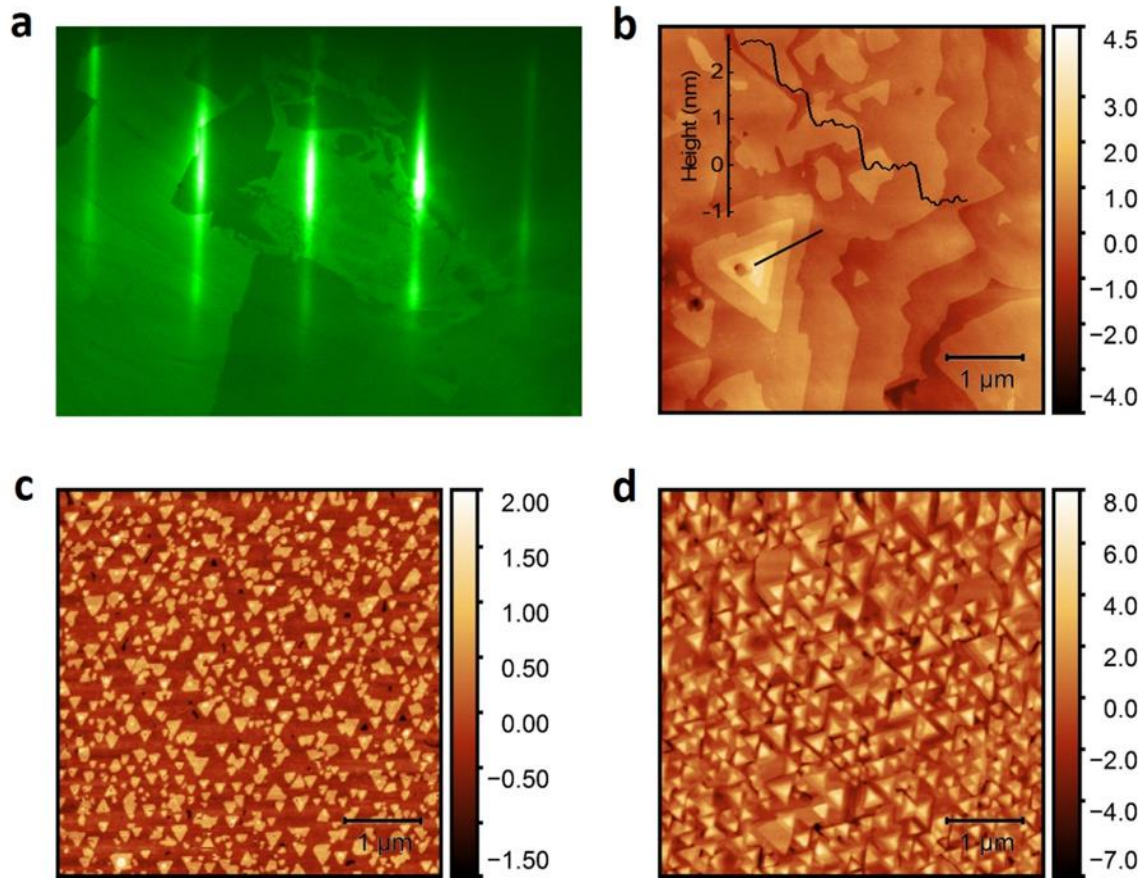

Figure 2: Characterizations of $Bi_2Se_3$ thin films grown on different substrates. (a) Sharp streaky RHEED pattern of the epitaxial $Bi_2Se_3$ thin film grown on $BiInSe_3$ substrate. (b) AFM image of a 10 QL $Bi_2Se_3$ thin film grown on $BiInSe_3$ substrate, exhibiting large area flat terraces. (c) AFM image of a 10 QL $Bi_2Se_3$ thin film directly grown on $Al_2O_3$. (d) AFM image of a 10 QL $Bi_2Se_3$ thin film grown on $In_2Se_3$-$BiInSe_3$ buffer layer using the method in ref 7. Units in all the AFM scale bars are nm.

Electrical transport property is another tool to evaluate the quality of a TI thin film, especially for characterizing the level of defects. We grew 20 QL $Bi_2Se_3$ on $BiInSe_3$ substrate with 100 nm Se capping on top, as shown in Fig. 3(a). Before investigating the transport properties of $Bi_2Se_3$ film, we rule out the shunting effect from the substrate by measuring the temperature dependent sheet resistance of the $BiInSe_3$ substrate alone, as shown in Fig. 3(b). Here the $BiInSe_3$ substrate was treated exactly the same way as the ones used for $Bi_2Se_3$ growth (mounted on $Al_2O_3$ substrate, cured with the adhesive, cleaved for a fresh surface and in-situ annealed at 600 ℃ in the MBE chamber). The sheet resistance of $BiInSe_3$ substrate shows typical insulating behavior, rising sharply with decreasing temperature. On the other hand, as shown in Fig. 3(c), the sheet resistance of $Bi_2Se_3$ grown on $BiInSe_3$ substrate exhibits typical metallic behavior [30]. The fact that $BiInSe_3$ is several orders more insulating than $Bi_2Se_3$ film at low temperature undoubtedly excludes the shunting effect from the $BiInSe_3$ substrate.

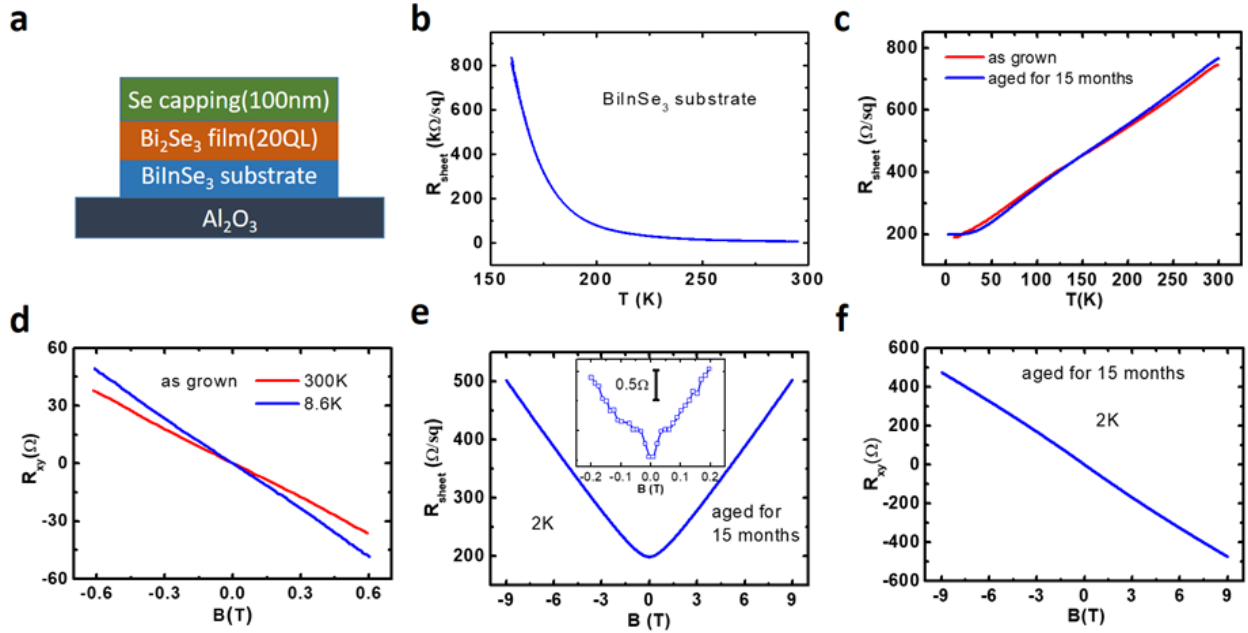

Figure 3: Transport properties of $Bi_2Se_3$ thin film grown on $BiInSe_3$ substrate. (a) Schematic illustration of the MBE growth for a $Bi_2Se_3$ thin film grown on $BiInSe_3$ substrate. The dummy $Al_2O_3$ substrate only acts as a holder because it fits well with our MBE sample plate. (b) The longitudinal sheet resistance for a $BiInSe_3$ substrate alone. (c) The longitudinal sheet resistance for a 20 QL $Bi_2Se_3$ thin film grown on $BiInSe_3$ substrate, measured in as grown state and aged state respectively. (d) The Hall resistance of the as grown $Bi_2Se_3$ sample measured at 300 K and 8.6 K. (e) The magnetoresistance of the same $Bi_2Se_3$ sample but aged for 15 months, measured at 2 K and up to 9 T. Inset shows the enlarged plot at low field range. (f) The Hall resistance of the same $Bi_2Se_3$ sample but aged for 15 months, measured at 2 K and up to 9 T.

Figure 3(d) shows the Hall effect data, which gives the 2D carrier density of $7.8 \times 10^{12}$ cm$^{-2}$ at 8.6 K. Combining the results of Hall effect and sheet resistance gives the Hall mobility of 4206 cm$^2$/V·s. The same $Bi_2Se_3$ film sample was sitting in air for 15 months and then remeasured again. As shown in Figure 3(c), the temperature dependent sheet resistance almost remains the same as the as-grown data. Figure 3(d) presents the magnetoresistance (MR) of the aged $Bi_2Se_3$ sample measured at 2 K and up to 9 T. The inset shows the enlarged plot at low field range, exhibiting an obvious resistance cusp feature, which is the typical signature of weak antilocalization (WAL). Figure 3(f) shows the Hall resistance of the aged $Bi_2Se_3$ sample, giving a higher 2D carrier density of $1.1 \times 10^{13}$ cm$^{-2}$ than the as-gown sample, which is the result of aging effect. The Shubnikov-de Haas (SdH) oscillation is absent in both MR and Hall resistance data, similar to the situation of $Bi_2Se_3$ films grown on $In_2Se_3$-$BiInSe_3$ buffer layer, even though they possess record-low 2D carrier density and highest mobility [7, 8]. In $Bi_2Se_3$, the presence of quantum oscillations is not directly related to the quality of films. Those $Bi_2Se_3$ films that exhibit quantum oscillations have high carrier densities typically on the order of $10^{13}$ $cm^{-2}$ with intermediate mobilities [31, 32]. Although the underlying reason for the absence of SdH oscillations in these low-carrier-density, high-mobility films still requires further exploration, one possibility is due to the carrier density inhomogeneity, which has been reported to significantly suppress the SdH oscillation in GaN/AlGaN heterostructures [33]. When the carrier density is low, the carrier density inhomogeneity could have more severe effect and result in the absence of SdH oscillation [34].

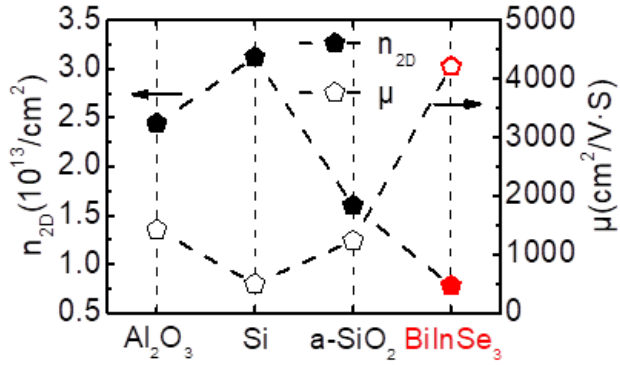

Figure 4: 2D carrier density and Hall mobility of 20 QL $Bi_2Se_3$ films grown on different substrates. Data for $Al_2O_3$, Si and a-$SiO_2$ were taken at 1.5 K, extracted from ref 22. Data for $BiInSe_3$ were taken at 8.6 K.

Figure 4 compares the 2D carrier density and Hall mobility of 20 QL $Bi_2Se_3$ films grown on various substrates: $BiInSe_3$ substrate clearly outperforms other substrates in both the residual carrier density and mobility. Among the substrates shown in Fig. 4, $BiInSe_3$ is the only one with the 2D carrier density below $10^{13}$ $cm^{-2}$ for 20 QL $Bi_2Se_3$ film. Meanwhile, the Hall mobility of Se capped $Bi_2Se_3$ grown on $BiInSe_3$ substrate achieves 4206 $cm^2/V·s$, significantly higher than on other substrates. It is worth mentioning that this value even outpaces the $In_2Se_3$-$BiInSe_3$ buffer layer scheme [7]. There could be multiple factors contributing to the superior transport properties of $Bi_2Se_3$ films grown on the $BiInSe_3$ substrate. First, the structural match should minimize structural defects at the interface. Second, the chemical match eliminates unintended doping effect at the interface [22]. Lastly, the much larger terraces imply correspondingly longer mean free path, leading to the larger Hall mobility.

## 3 Conclusion

In recent years, TI thin films have been extensively investigated for both fundamental studies and spintronic applications such as spin-orbit torque devices [35-38]. Nonetheless, the existing commercial substrates lead to large residual carrier densities and low mobilities and hamper the progress of TI research. The newly introduced $BiInSe_3$ substrate, which is ideally matched both structurally and chemically with $Bi_2Se_3$ films and can easily achieve atomically flat surfaces, will accelerate the advance of TI studies beyond the current limit.

## 4 Method

**Transport measurement**: The resistance and Hall resistance measurements were performed with the standard van der Pauw geometry in both a closed-cycle cryostat (8.6 K, 0.6 T) and a Quantum Design Physical Property Measurement System (PPMS; 2K, 9T). Electrical electrodes were made by manually pressing four indium wires on the corners of each sample. All the samples were carefully cut into square shape to minimize the deviation from van der Pauw geometry. Raw data of $R_{xx}$ and $R_{xy}$ were properly symmetrized and antisymmetrized respectively. 2D carrier density was extracted from $n_{2D}=1/e(dR_{xy}/dB)^{-1}$ where e is the electronic charge and $dR_{xy}/dB$ is the slope of the Hall resistance vs magnetic field B, measured at the origin. The zero-field sheet resistance was calculated from $R_{sheet}=R_{xx}(B=0) \pi/\ln(2)$ for the van der Pauw geometry. Hall mobility μ was calculated by using $μ= (eR_{sheet} n_{2D})^{-1}$.

## Acknowledgements

This work is supported by the center for Quantum Materials Synthesis (cQMS), funded by the Gordon and Betty Moore Foundation's EPiQS initiative through grant GBMF6402, and by Rutgers University.
## References

[1] Chang, C. Z.; Zhang, J.; Feng, X.; Shen, J.; Zhang, Z.; Guo, M.; Li, K.; Ou, Y.; Wei, P.; Wang, L. L. et al. Experimental observation of the quantum anomalous Hall effect in a magnetic topological insulator. *Science* **2013,** *340*, 167-170.

[2] Checkelsky, J. G.; Yoshimi, R.; Tsukazaki, A.; Takahashi, K. S.; Kozuka, Y.; Falson, J.; Kawasaki, M.; Tokura, Y. Trajectory of the anomalous Hall effect towards the quantized state in a ferromagnetic topological insulator. *Nat. Phys.* **2014,** *10*, 731-736.

[3] Kou, X.; Guo, S. T.; Fan, Y.; Pan, L.; Lang, M.; Jiang, Y.; Shao, Q.; Nie, T.; Murata, K.; Tang, J. et al. Scale-invariant quantum anomalous Hall effect in magnetic topological insulators beyond the two-dimensional limit. *Phys. Rev. Lett.* **2014,** *113*, 137201.

[4] Yoshimi, R.; Tsukazaki, A.; Kozuka, Y.; Falson, J.; Takahashi, K. S.; Checkelsky, J. G.; Nagaosa, N.; Kawasaki, M.; Tokura, Y. Quantum Hall effect on top and bottom surface states of topological insulator $(Bi_{1-x}Sb_x)_2Te_3$ films. *Nat. Commun.* **2015,** *6*, 6627.

[5] Xu, Y.; Miotkowski, I.; Liu, C.; Tian, J.; Nam, H.; Alidoust, N.; Hu, J.; Shih, C.-K.; Hasan, M. Z.; Chen, Y. P. Observation of topological surface state quantum Hall effect in an intrinsic three-dimensional topological insulator. *Nat. Phys.* **2014,** *10*, 956-963.

[6] Xu, Y.; Miotkowski, I.; Chen, Y. P. Quantum transport of two-species Dirac fermions in dual-gated three-dimensional topological insulators. *Nat. Commun.* **2016,** *7*, 11434.

[7] Koirala, N.; Brahlek, M.; Salehi, M.; Wu, L.; Dai, J.; Waugh, J.; Nummy, T.; Han, M. G.; Moon, J.; Zhu, Y. et al. Record Surface State Mobility and Quantum Hall Effect in Topological Insulator Thin Films via Interface Engineering. *Nano Lett.* **2015,** *15*, 8245-8249.

[8] Moon, J.; Koirala, N.; Salehi, M.; Zhang, W.; Wu, W.; Oh, S. Solution to the Hole-Doping Problem and Tunable Quantum Hall Effect in $Bi_2Se_3$ Thin Films. *Nano Lett.* **2018,** *18*, 820-826.

[9] Salehi, M.; Shapourian, H.; Rosen, I. T.; Han, M.-G.; Moon, J.; Shibayev, P.; Jain, D.; Goldhaber-Gordon, D.; Oh, S. Quantum-Hall to Insulator Transition in Ultra-Low-Carrier-Density Topological Insulator Films and a Hidden Phase of the Zeroth Landau Level. *Adv. Mater.* **2019,** 1901091.

[10] Koirala, N.; Salehi, M.; Moon, J.; Oh, S. Gate-tunable quantum Hall effects in defect-suppressed $Bi_2Se_3$ films. *Phys. Rev. B* **2019,** *100*, 085404.

[11] Fei, F.; Zhang, S.; Zhang, M.; Shah, S. A.; Song, F.; Wang, X.; Wang, B. The Material Efforts for Quantized Hall Devices Based on Topological Insulators. *Adv. Mater.* **2019,** 1904593.

[12] Mogi, M.; Kawamura, M.; Tsukazaki, A.; Yoshimi, R.; Takahashi, K. S.; Kawasaki, M.; Tokura, Y. Tailoring tricolor structure of magnetic topological insulator for robust axion insulator. *Sci. Adv.* **2017,** *3*, eaao1669.

[13] Mogi, M.; Kawamura, M.; Yoshimi, R.; Tsukazaki, A.; Kozuka, Y.; Shirakawa, N.; Takahashi, K. S.; Kawasaki, M.; Tokura, Y. A magnetic heterostructure of topological insulators as a candidate for an axion insulator. *Nat. Mater.* **2017,** *16*, 516-521.

[14] Xiao, D.; Jiang, J.; Shin, J. H.; Wang, W.; Wang, F.; Zhao, Y. F.; Liu, C.; Wu, W.; Chan, M. H. W.; Samarth, N. et al. Realization of the Axion Insulator State in Quantum Anomalous Hall Sandwich Heterostructures. *Phys. Rev. Lett.* **2018,** *120*, 056801.

[15] Wu, L.; Salehi, M.; Koirala, N.; Moon, J.; Oh, S.; Armitage, N. P. Quantized Faraday and Kerr rotation and axion electrodynamics of a 3D topological insulator. *Science* **2016,** *354*, 1124-1127.

[16] Yoshida, S.; Misawa, S.; Gonda, S. Improvements on the electrical and luminescent properties of reactive molecular beam epitaxially grown GaN films by using AlN‐coated sapphire substrates. *Appl. Phys. Lett.* **1983,** *42*, 427-429.

[17] Amano, H.; Sawaki, N.; Akasaki, I.; Toyoda, Y. Metalorganic vapor phase epitaxial growth of a high quality GaN film using an AlN buffer layer. *Appl. Phys. Lett.* **1986,** *48*, 353-355.

[18] Nakamura, S. GaN Growth Using GaN Buffer Layer. *Jpn. J. Appl. Phys.* **1991,** *30*, L1705-L1707.

[19] Li, G.; Wang, W.; Yang, W.; Lin, Y.; Wang, H.; Lin, Z.; Zhou, S. GaN-based light-emitting diodes on various substrates: a critical review. *Rep. Prog. Phys.* **2016,** *79*, 056501.

[20] Tang, C.; Chang, C. Z.; Zhao, G.; Liu, Y.; Jiang, Z.; Liu, C. X.; McCartney, M. R.; Smith, D. J.; Chen, T.; Moodera, J. S. et al. Above 400-K robust perpendicular ferromagnetic phase in a topological insulator. *Sci. Adv.* **2017,** *3*, e1700307.

[21] Yao, X.; Gao, B.; Han, M. G.; Jain, D.; Moon, J.; Kim, J. W.; Zhu, Y.; Cheong, S. W.; Oh, S. Record High-Proximity-Induced Anomalous Hall Effect in $(Bi_xSb_{1-x})_2Te_3$ Thin Film Grown on $CrGeTe_3$ Substrate. *Nano Lett.* **2019,** *19*, 4567-4573.

[22] Bansal, N.; Koirala, N.; Brahlek, M.; Han, M.-G.; Zhu, Y.; Cao, Y.; Waugh, J.; Dessau, D. S.; Oh, S. Robust topological surface states of $Bi_2Se_3$ thin films on amorphous $SiO_2$/Si substrate and a large ambipolar gating effect. *Appl. Phys. Lett.* **2014,** *104*, 241606.

[23] Guo, X.; Xu, Z. J.; Liu, H. C.; Zhao, B.; Dai, X. Q.; He, H. T.; Wang, J. N.; Liu, H. J.; Ho, W. K.; Xie, M. H. Single domain $Bi_2Se_3$ films grown on InP(111)A by molecular-beam epitaxy. *Appl. Phys. Lett.* **2013,** *102*, 151604.
7